\documentclass[
    aps,
    prl,
    english,
    twocolumn,
    showpacs,
    preprintnumbers,
    amsmath,
    amssymb,
    floatfix,
    superscriptaddress,
    longbibliography
]{revtex4-2}

\usepackage[T1]{fontenc}
\usepackage[latin9]{inputenc}
\usepackage{graphicx}
\usepackage{amssymb}
\usepackage{babel}

\global\arraycolsep=2pt
\usepackage{amsmath}
\usepackage{amssymb}
\usepackage{graphicx}
\usepackage{textcomp}
\usepackage{calrsfs}
\usepackage{caption}
\usepackage{subcaption}
\usepackage{bm}
\usepackage{color}
\usepackage{geometry}
\usepackage{multirow}
\usepackage{array}

\newcommand{\ben}{\begin{equation*}}
\newcommand{\een}{\end{equation*}}
\newcommand{\bean}{\begin{eqnarray*}}
\newcommand{\eean}{\end{eqnarray*}}

\newcommand{\be}{\begin{equation}}
\newcommand{\ee}{\end{equation}}
\newcommand{\bea}{\begin{eqnarray}}
\newcommand{\eea}{\end{eqnarray}}
\usepackage{orcidlink}

\usepackage{upgreek}

\usepackage[version=3]{mhchem}

\usepackage{color}
\usepackage[mathscr]{euscript}

\usepackage{hyperref}
\hypersetup{
    colorlinks=true,
    urlcolor  = blue,
    citecolor=magenta, % cite
    linkcolor=magenta % ref
}

\makeatother
\usepackage{babel}

\begin{document}
    %\title[Repulsion-Attraction Casimir Force Transitions Between Cylindrical  Nanoparticles in Magnetic Fluid]{Repulsion-Attraction Casimir Force Transitions Between Cylindrical  Nanoparticles in Magnetic Fluid}

    %\title{Trapping of Nanoparticles Controlled via Magnetic Casimir Force}
    %\title{Trapping Distances in a Force Set-Up Controlled by Solution Permeability}
    \title{Trapping in a Casimir Force Set-Up Controlled by Solution
    Permeability}

    \author{S. Pal\,\orcidlink{0009-0003-6356-3409}\,}
    \affiliation{Dipartimento di Fisica e Chimica-Emilio Segr\`e, \href{https://www.unipa.it/target/international-students/en/about/the-university/}{Universit\`a degli Studi di Palermo}, Via Archirafi 36, 90123 Palermo, Italy}

    \author{L. Inacio\,\orcidlink{0000-0001-8971-0591}\,} % \orcidlink{}\,
    \affiliation{\href{https://ensemble3.eu/}{Centre of Excellence ENSEMBLE3} Sp. z o. o., Wolczynska Str. 133, 01-919, Warsaw, Poland}

    \author{L. M. Woods \orcidlink{0000-0002-9872-1847}\,}
    \affiliation{Department of Physics, \href{https://www.usf.edu}{University of South Florida}, Tampa, FL, 33620, USA}

    \author{U. De Giovannini \orcidlink{0000-0002-4899-1304}\,}
    \affiliation{Dipartimento di Fisica e Chimica-Emilio Segr\`e, \href{https://www.unipa.it/target/international-students/en/about/the-university/}{Universit\`a degli Studi di Palermo}, Via Archirafi 36, 90123 Palermo, Italy}
    \affiliation{\href{https://www.mpsd.mpg.de/person/44457/2736}{Max Planck Institute for the Structure and Dynamics of Matter}, Luruper Chaussee 149, Hamburg, Germany}
    \author{C. Persson \orcidlink{0000-0002-9050-5445}\,}
    \affiliation{Department of Materials Science and Engineering, \href{https://www.kth.se/en}{KTH Royal Institute of Technology}, SE-100 44 Stockholm, Sweden}

    \author{M. Bostr{\"o}m \orcidlink{0000-0001-5111-4049}\,}
    \email{mathias.bostrom@ensemble3.eu} \affiliation{\href{https://ensemble3.eu/}{Centre of Excellence ENSEMBLE3} Sp. z o. o., Wolczynska Str. 133, 01-919, Warsaw, Poland}
    \affiliation{Chemical and Biological Systems Simulation Lab, \href{https://cent.uw.edu.pl/}{Centre of New Technologies, University of Warsaw}, Banacha 2C, 02-097 Warsaw, Poland}
    \date{\today}

    \begin{abstract}
        {%It is essential to reduce attraction from short-range forces causing clustering of building blocks in nanoengineering. 
        We have designed a system enabling tuning Casimir attraction/repulsion transitions between a polystyrene sphere attached to an atomic force microscope tip near a Teflon surface in a magnetic fluid mixture. Notably, the trapping distances can be changed by several orders of magnitude by changes in zero-frequency transverse electric contributions to the Casimir force. We demonstrate that this can be achieved via modifications in the average diameter for the magnetite particles while keeping the magnetite volume fractions fixed. }
    \end{abstract}

    \maketitle
% \textcolor{red}{I think the notation in figures, formulas, text needs synchronization: consider writing the abbreviation of the materials on Fig. 1; $\phi$ on Fig. 2 is not defined, maybe substitute-interchange with concentration $p$ in text}

% \textcolor{red}{Fig. 2 needs some description, especially the disparity between PS and the others at low frequency}

% \textcolor{red}{explicit $\xi_m$ is needed in the formulas}

% \textcolor{red}{Somewhere the use of this effective model for magnetic and optical properties needs justification}

\vspace{0.5cm}

    The Casimir-Lifshitz theories have in recent years been intensely explored, in particular for ubiquitous interactions between objects. Numerous investigations have led to many exciting results showing new ways of probing basic physics of materials through quantum vacuum excitations \cite{Klimchitskaya2009,WoodsRevModPhys.88.045003}. Does    this imply that there is little hope for further significant advancements?    A fundamental question worthy of exploration is if it is possible to    manipulate the force that controls the physics of the nanoscale world not     using exotic materials or extreme conditions, but through a clever selection  of ordinary substances. Here we show that by exploiting the material properties, especially the    {\it magnetic susceptibility of liquids}, we could unveil new ways of Casimir interaction tuning in a    sphere-plane geometry through permeability controlled trapping of    polystyrene (PS) particles near Teflon (PTFE) surface. 
    
\begin{figure}[!ht]
        \centering
        \includegraphics[width=\columnwidth]{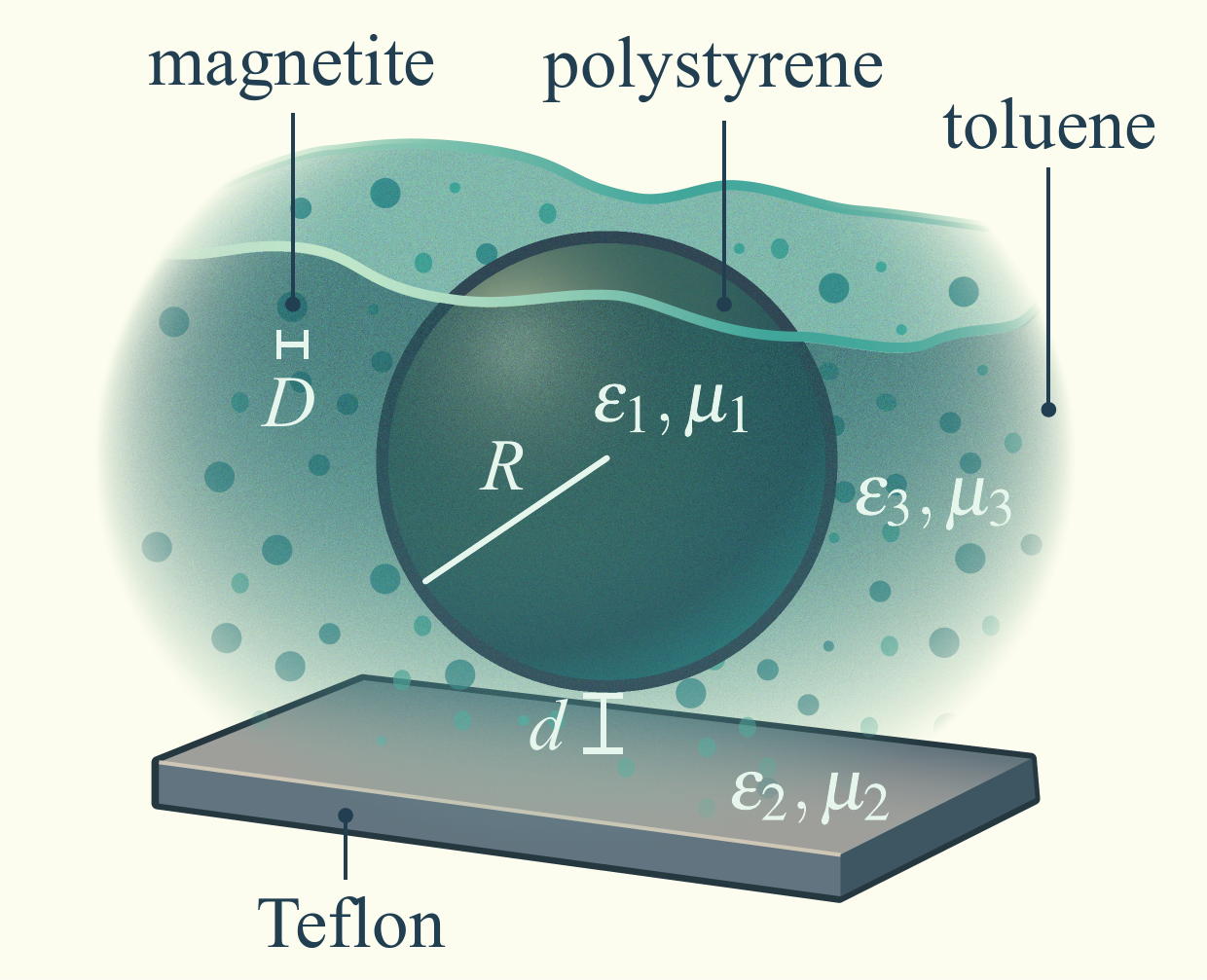}
        \caption{(Colors online) Schematics of the force set-up
        system. The dielectric permittivity, $\varepsilon_i$, and magnetic susceptibility, $\mu_i$, functions denote the response properties for a polystyrene (PS) sphere with radius $R$, Teflon (PTFE) substrate , and toluene fluid with dispersed magnetite nano-particles with diameter $D$. The sphere-substrate distance is $d$.}
        \label{fig:schematic}
    \end{figure}

    A schematic illustration of the system configuration of interest here is shown in Fig.\,(\ref{fig:schematic}).  Intriguingly for a Casimir force setup immersed in a fluid medium~\cite{DuckerSendenPashley_AFM_Nature,Zwol1}, both the separation at which the interaction reaches a stable equilibrium between objects, here referred to as {\it trapping distance}, and the strength of the repulsion and attraction, can be tuned by several orders of magnitude simply by modulating the magnetic permeability of the fluid. Notably, changes in the magnetic response do not affect the optical properties of the fluid significantly. Such tuning results in a rare and remarkable situation where the Casimir force is dominated by {\it transverse electric (TE) modes} as opposed to the conventional situation, in which the interaction is primarily determined by transverse magnetic (TM) contributions. 
    %It is observed that trapping distances in a Casimir force set-up within a fluid medium\,\cite{DuckerSendenPashley_AFM_Nature,Zwol1}, and the magnitudes of both repulsion and attraction can be tuned orders of magnitudes via subtle changes in magnetic fluid permeability alone.
    %Notably, we keep the optical properties of the fluid largely unchanged. It turns out that cancellations within the usually dominating transverse magnetic contributions enable a unique case where the forces are predicted to be dominated by transverse electric modes.
    Our interest in TE modes dates back 25 years\,\cite{Bost2000} in the context of understanding how dissipation influences contributions from zero frequency TE and TM modes in Casimir force measurements between metallic surfaces. The TE modes are especially prominent when thermal effects are taken into account and the experiments reported in Ref.\,\cite{Lamo1997,Lamo1998} had already shown that it is possible to directly access Casimir forces with unprecedented precision in various regimes. Since then, it has become a prominent research field in Casimir related nanotechnology by finding ways to change the sign of the interaction through tailoring the materials optical properties\,\cite{Zwol1,WoodsRevModPhys.88.045003,Woods2020,MundayTorque2022,shelden2023enhanced,MundayReview2024,Woods2024}. Such research encompasses investigating materials combinations and geometries that allow the manipulation of dispersion forces at the nanoscale level aiming to maximize functionality in micro and nanoscale devices.
 %   Our interest in the these modes dates back 25 years\,\cite{Bost2000} when we  explored how dissipation influences contributions from zero frequency transverse    electric mode in the Casimir force measurements between metallic surfaces. At that time the high-precision experiments by Lamoreaux\,\cite{Lamo1997,Lamo1998}    had just shown that it was possible to measure Casimir forces with high   accuracy. Much focus in Casimir-related nanotechnology has been on controlling   attractive and repulsive forces via selecting materials with certain optical    properties\,\cite{Zwol1,WoodsRevModPhys.88.045003,Woods2020,MundayTorque2022,MundayReview2024,Woods2024}.

The ultimate target in our study is to effectively control the trapping of nanoparticles via repulsion-attraction force transitions by influencing TE mode contributions through the magnetic susceptibility of the immersing fluid. Repulsion, in particular, was found for thin films
    of liquid helium absorbed on different fluoride surfaces \cite{AndSab}. The film thicknesses, ranging from 10 to 200\,\AA \, %\cite{AndSab},
    could be measured to within a few percent in most cases. %For these  saturated-film measurements the repulsive van der Waals potential was equal to the negative of the gravitational potential \textcolor{red}{this sentence may be reomved; it does not provide any significance to the paper in my view} \,\cite{AndSab}. 
    A good agreement 
    between these experimental data and the results from Lifshitz theory was reported \cite{Rich71}. Other Casimir force experiments, also modeled by the Casimir-Lifshitz theory, report accurate measurements of thickness hydrocarbon films on water near the alkane saturated vapor pressure  \,\cite{Haux}. 
Attempts to go beyond material combinations with pure attraction or pure repulsion\,\cite{MILLING1996460,Lee2002,Feiler2008,Munday2009,Zwol1} have also been of interest. Systems with attraction-repulsion transitions were, for example, found with topological insulators\,\cite{PhysRevLett.106.020403,PhysRevB.88.085421},  tin\,\cite{PhysRevB.97.125421}, and vanadium dioxide\,\cite{PhysRevB.101.104107,GE2022128392}. While in most systems, there are no stable equilibria, Zhao {\it et al.}\,\cite{zhao2019stable} experimentally showed Casimir trapping of objects near substrates. Twist-induced attraction-repulsion crossover between layers of lithium iodate interacting across a nanofluid has also been demonstrated\,\cite{twistinducedCasimirswitch2024}.

Additional Casimir force flexibility can be engineered using magnetic fluids. The combination of magnetic permeability\,\cite{klimchitskaya2019impact,VelichkoKlimchitskaya2020} and
magnetic field\,\cite{ZhangNature2024,MoazzamiNature2025} was in the past considered to tune the interaction. 
In particular, Zhang {\it et al.}\,\cite{ZhangNature2024} proposed that the permeability of an intermediate mixture consisting of a fluid and magnetic nanoparticles approach that of a purely dielectric media already at low magnetic fields.
In our work, a mechanism of tuning magnetic Casimir trapping is also revealed. We show a reliable method to engineer the characteristics of fluid media for the 
    control of repulsion-attraction transitions via magnetic permeability changes
    while keeping optical properties largely unaltered. Notably, surprising impacts from retardation and thermal effects are predicted at distances
    ranging from many hundreds of nanometers down to almost contact distances. The essence of the approach
    consists of modulating the fluid magnetic permeability  via clever choices of volume fractions and particle sizes of the immersed magnetite nanoparticles in a toluene solution.

    The system we are interested in is an atomic force microscope (AFM)
    with a micronsized polystyrene sphere attached to an AFM tip near a Teflon
    surface illustrated in Fig.\,\ref{fig:schematic}. The sphere-plate interaction is within the toluene fluid containing a mixture of randomly dispersed magnetite
    nanoparticles. Based on Lifshitz theory,  at separations ($d$) much
    smaller than the radius of the sphere (with radius $R$\,$\sim$1--50$\,\mu$m)
    the Casimir force at temperature $T$ can be written as a sum over Matsubara frequencies $\xi_{\rm m}=m\, 2\pi k_{\rm B}T/\hbar$ ($\hbar$ - Plank's constant, $k_{\rm B}$ -  Boltzmann constant) ~\cite{Dzya,NinhamParsegianWeiss1970,Richmond_1971}
    \begin{equation}
        \begin{aligned}
            \frac{f(d)}{R}=k_{\rm B}T{\sum_{m=0}^\infty}{}^{\prime}\int\limits_{0}^{\infty}d k^{\parallel}\, k^{\parallel} \hspace{-11pt}\sum_{\sigma=\rm {TE, TM}}\hspace{-10pt}\ln\hspace{-1pt}\Big[1 -\\r_{\sigma}^{31}(\xi_m)r_{\sigma}^{32}(\xi_m)\mathrm{e}^{-2\kappa_3^\perp(\xi_m) d}\Big] . \label{forcedivR}
        \end{aligned}
    \end{equation}    
    
    Here, ${k}^{\parallel}$ is the component of the wave vector ${\bf k}=(2\pi/\lambda)\bf{\hat{k}}$ parallel to the surface, and $\kappa_{i}^{\perp}(\xi_m)= {\scriptstyle \sqrt{{k^\parallel}^{2}+\mu_{i}\varepsilon_{i}\mathrm{\xi}_{m}^{2}/c^{2}}}$ with $c$ being the speed of light, $\varepsilon_{i}$ - the dielectric function, $\mu_{i}$ - the magnetic permeability with indices $i$ for the sphere ($i=$1), plate ($i=$2), and fluid ($i=$3).
    The primed sum in the Eq. (\ref{forcedivR}) denotes that the zeroth term ($m$ = 0) is weighted by $1/2$.  The Fresnel reflection coefficients for the TE and TM polarizations, $r_{\mathrm{TE},\mathrm{TM}}^{ij}$, by which a photon in medium $i$ is reflected from medium $j$ are\,\cite{Richmond_1971}
    \begin{equation}
        r_{\rm TE}^{ij}= \frac{\mu_{j}\kappa^{\perp}_{i}-\mu_{i}\kappa^{\perp}_{j}}{\mu_{j}\kappa^{\perp}_{i}+\mu_{i}\kappa^{\perp}_{j}}
        \,, \hspace{-3pt}\quad r_{\rm TM}^{ij}= \frac{\varepsilon_{j}\kappa^{\perp}_{i}-\varepsilon_{i}\kappa^{\perp}_{j}}{\varepsilon_{j}\kappa^{\perp}_{i}+\varepsilon_{i}\kappa^{\perp}_{j}}
        \,. \label{eq:rtTETM}
    \end{equation}

From Eqs. (\ref{forcedivR},\ref{eq:rtTETM}) one finds the zero-frequency TE and TM asymptotes, 

\begin{align}
    \frac{f(d)_{\sigma}^{m=0}}{R}=\frac{k_{\rm B}T}{8 d^2}\int\limits_{0}^{\infty}d x \, x \ln{\scriptstyle\left[1-( r_{\sigma}^{31} r_{\sigma}^{32}){\big|}_{\xi=0}\,\mathrm{e}^{-x} \right]}
\end{align}
with  ${\scriptstyle r_{\rm TE}^{ij}{\big|}_{\xi=0}=\frac{\mu_{j}(0)-\mu_{i}(0)}{\mu_{j}(0)+\mu_{i}(0)}}$ and ${\scriptstyle r_{\rm TM}^{ij}{\big|}_{\xi=0}=  \frac{\varepsilon_{j}(0)-\varepsilon_{i}(0)}{\varepsilon_{j}(0)+\varepsilon_{i}(0)}}$. %\textcolor{red}{I think the following equations can be moved here in the text, and not given as separate formulas}
%\begin{equation}
 %       r_{\rm TE}^{ij}{\bigg|}_{\xi=0}= \frac{\mu_{j}(0)-\mu_{i}(0)}{\mu_{j}(0)+\mu_{i}(0)}\,
   %     , \hspace{-3pt}\quad r_{\rm TM}^{ij}{\bigg|}_{\xi=0}= \frac{\varepsilon_{j}(0)-\varepsilon_{i}(0)}{\varepsilon_{j}(0)+\varepsilon_{i}(0)}
 %       \,. \label{eq:rtTETMzerofreq}
%    \end{equation}
We would like to engineer a system where the effect of {{\it zero-frequency TE mode}} can be measured and in doing so we provide a new pathway to control
    trapping at the nano and micro scale. This contribution was the source of the so-called
    Drude-Plasma controversy\,\cite{Bost2000} in connection with Casimir force
    measurements between widely separated metal surfaces\,\cite{Lamo1997,Lamo1998,HarrisPhysRevA.62.052109,RevModPhys.81.1827,SushNP,Mohid,Ser2018}. 
   It appears that different models for the optical response of the metals lead to different results of the force asymptote at large distances. Since the zero-frequency TE modes play a major role at large separations, as evident from Lifshitz theory, it is important to understand how it can be explored in magnetic systems. In some exotic cases,
    demonstrated here, the repulsion-attraction transition can happen at widely different separations, even at the nanoscale.
    \begin{figure}[!ht]
        \centering
        \includegraphics[width=0.85\columnwidth]{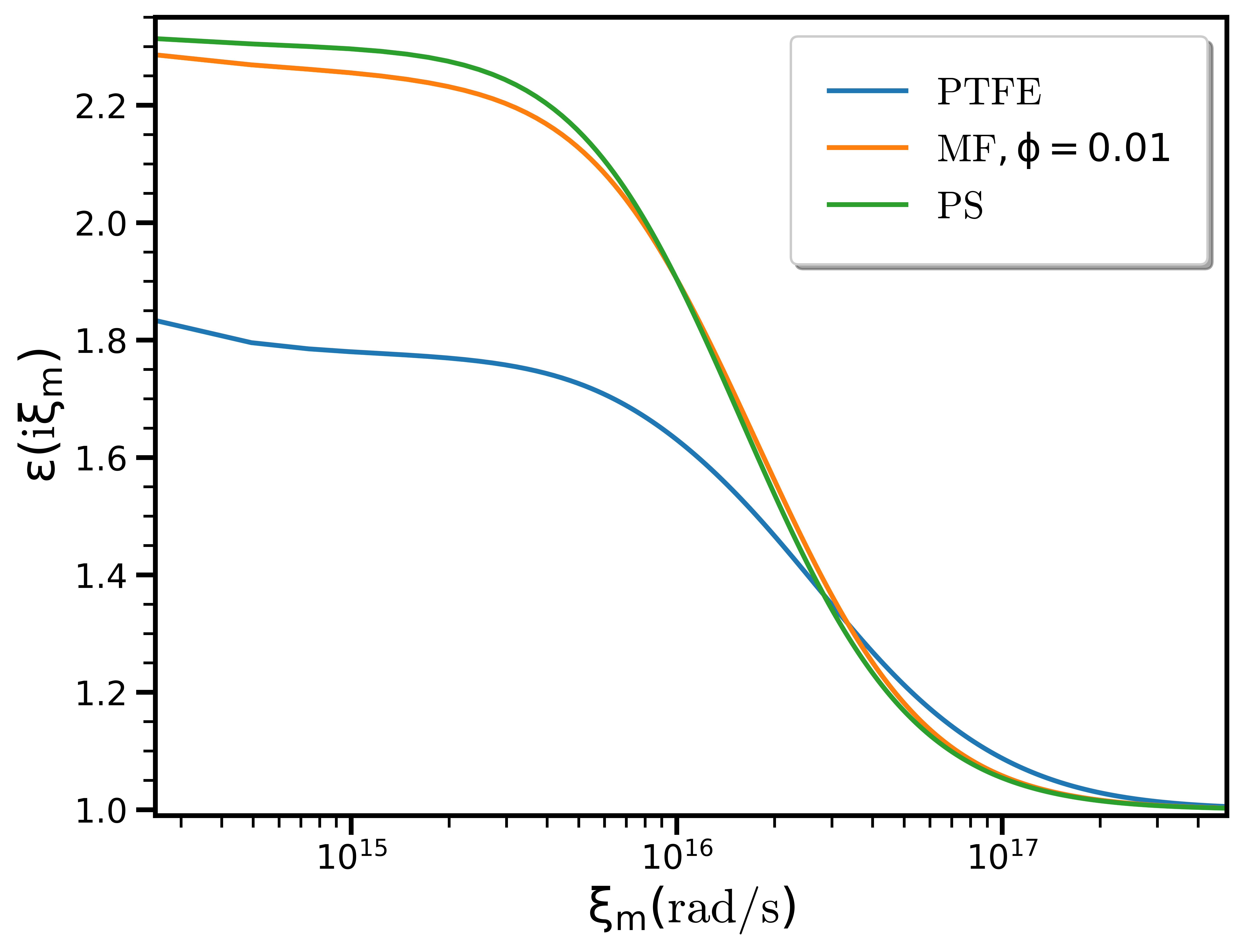}
        \vspace{-0.1in}
        \caption{(Colors online) The parametrized dielectric functions, as a function of imaginary frequencies
        ${\xi}_{m}$, for  Teflon (PTFE), polystyrene (PS), and toluene based on optical experiments given by van Zwol {\it et al.}\,\cite{Zwol1}. 
        The dielectric function for the magnetic fluid (MF) is calculated using  Eq.\,(\ref{3mixturediel}) and data from  \,\cite{PhysRevB.111.085407}. %\textcolor{red}{I think the next sentence can be removed}.       For the magnetic fluid (MF), we combine (using volume fraction $\Phi$ of magnetite and Eq.\,(\ref{3mixturediel})) the dielectric functions for toluene and magnetite\,\cite{PhysRevB.111.085407}.
        } 
        \label{fig:dielectric}
    \end{figure}

 For the system in Fig.\,\ref{fig:schematic}, we need to resolve the response properties of the involved materials. Here, we use the 
    parameterized dielectric function of magnetite as described recently\,\cite{PhysRevB.111.085407}.
    We use experimental data from the literature for toluene, Teflon, and polystyrene\,\cite{Zwol1}
    valid around room temperature (for calculations of forces we apply these
    data for $T = 298.15$\,K). To describe the effect of the volume fraction of
    magnetite nanoparticles we use a Lorentz-Lorenz-like model for the effective
    dielectric function\,\cite{Aspnes,Krag_LorenzLorentzFormula},
    \begin{equation}
        \varepsilon_{3}=\frac{1+2 \Lambda_{3}}{1-\Lambda_{3}}; \quad \Lambda_{3}=
        \sum_{\nu=1,2}\Phi_{\nu}\frac{\varepsilon_{3,\nu}-1}{\varepsilon_{3,\nu}+2}
        , \label{3mixturediel}
    \end{equation}
    where $\Phi_{\nu}$ and $\varepsilon_{3,\nu}$ are the volume fraction and dielectric
    function of the toluene ($\nu=1$) and magnetite ($\nu=2$). In the current
    work, the magnetite concentration is varied in the range $\Phi=0-2$\%  with changing diameter $D=2-20$\,nm. 
%    The zero frequency dielectric constants can be derived from the references (e.g., $\varepsilon_\mathrm{PTFE}(0)=2.1$, $\varepsilon_\mathrm{PS}(0)=2.4$, and $\varepsilon_\mathrm{MF,0.01}(0)=2.8$).
    Both Teflon surface and polystyrene are non-magnetic ($\mu_{1}(0)\approx\mu_{2}
    (0)\approx1$). 
    %The magnetic permeability of magnetite can be described using a Lorentz model with a magnetic susceptibility 
    %\textcolor{red}{this sentence is confusing:  Lorentz vs Lorentz-Lorentz-like model?} . % ($\chi$).  
   % \begin{equation}
   %     \mu(i\xi_{m})\sim 1+\frac{4\pi \chi}{1+\xi_{m}^{2}/\omega_{0}^{2}}.
   % \end{equation}
    Since the magnetic relaxation frequency ($\sim10^{10}$\,rad/s or less\,\cite{NylandBrevik1994}) is very low compared to the first Matsubara frequency, only the zero frequency contribution is significantly
    influenced by the magnetic effects. The effective zero frequency magnetic permeability for a water+magnetite
 fluid mixture can be modelled as\,\cite{klimchitskaya2019impact,VelichkoKlimchitskaya2020},
    \begin{equation}
        \mu_{\mathrm{mix}} \equiv \mu_{3}(0)=1+\frac{2\pi^{2}\Phi}{9}\frac{M_{s}^{2}D^{3}}{k_{B}T}, \label{mu3}
    \end{equation}
   where $M_{s}$ is the saturation magnetization per unit volume. Toluene is close to nonmagnetic
    ($\mu=0.99996$) so within the level of accuracy in our work we assume
    that the permeability of toluene-based mixtures can be estimated with the approximation
    used for water-based mixtures\,\cite{klimchitskaya2019impact,VelichkoKlimchitskaya2020}.  
  Examples of dielectric functions as a function of Matsubara frequencies are presented in Fig.\,(\ref{fig:dielectric}). 
  {The closeness of the dielectric functions for polystyrene and the magnetic fluid is essential to have a system where the transverse magnetic contribution largely cancels out.}
 % \textcolor{red}{some mention about the behavior - in low frequency PTFE has smaller plateau while MF and PS are almost identical. Is this important? Does this change if you change $\Phi$. Maybe you can plot a few curves for different Phi? }
    
       \begin{figure}[!ht]
        \centering
        \includegraphics[width=0.9\columnwidth]{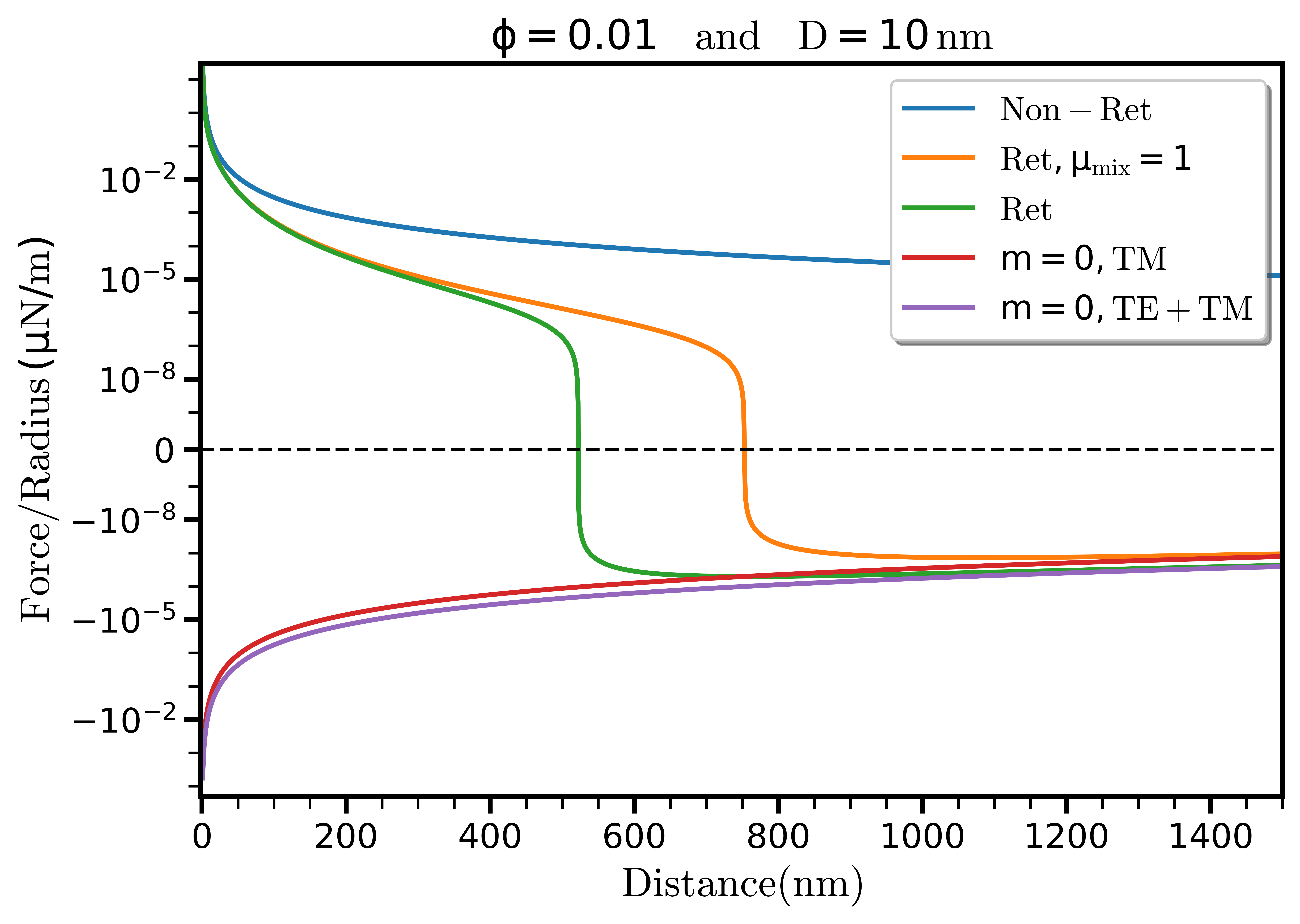}
        \caption{(Colors online) 
        The Casimir-Lifshitz force normalized by the radius of the PS sphere calculated using Eq.\,\ref{forcedivR} for the fully retarded (Ret), nonretarded limit (Non-Ret) and  with $\mu_{\mathrm{mix}}=1$. The zero Matsubara contributions for the TM (m=0, TM) and the TE and TM modes (m=0, TE+TM) are also shown. Here $\Phi=0.01$, $D=10$\,nm, and $T$=298.15\,K.}% Different contributions for the force normalized with radius of the PS sphere for fixed $\Phi=0.01$, magnetite diameter 10\,nm, and $T$=298.15\,K. We present the fully retarded (Ret) curves  with and without the permeability being different from one, the non-retarded (Non-Ret) curve, and the zero frequency contributions to the two retarded cases.}
        \label{fig:forcedivRadie_different_different_contributions}
    \end{figure}
    
        \begin{figure*}[!ht]
         \centering
        \includegraphics[width=0.9\textwidth,height=5.5cm]{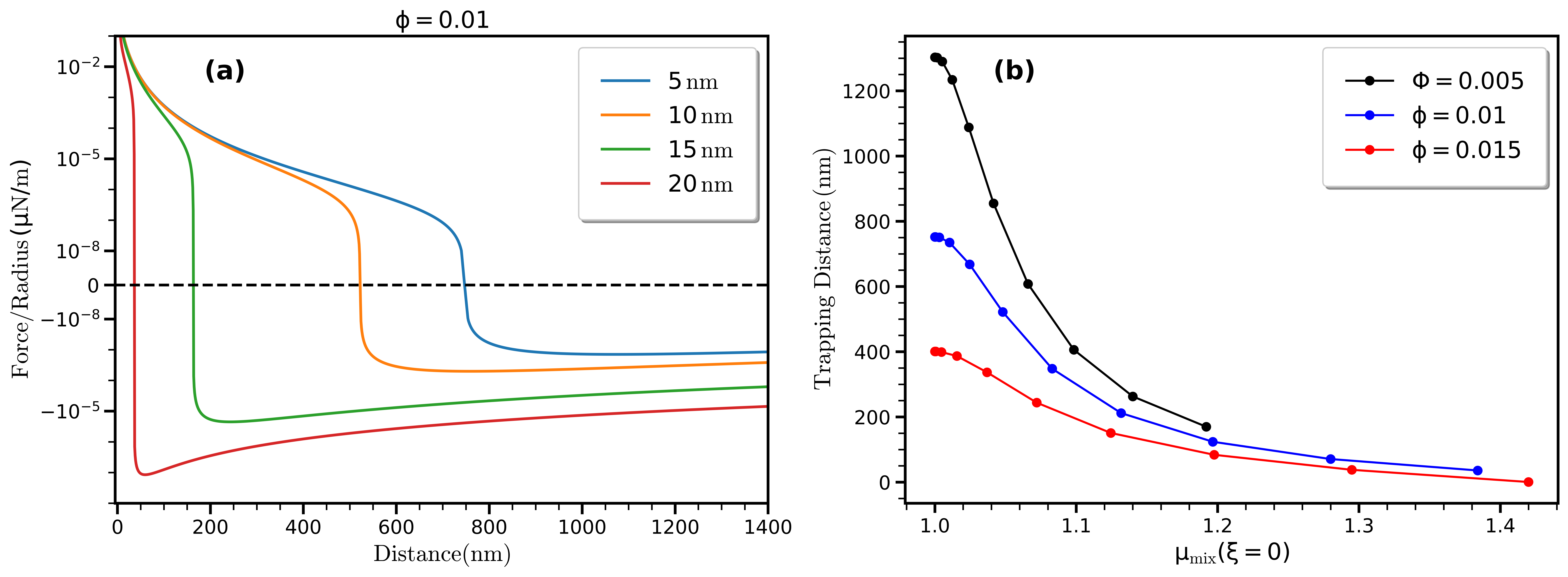}
        \caption{(Colors online) (a) Force normalized with radius of the PS sphere
        for a magnetite volume fraction $\Phi=0.01$ and and $T$=298.15\,K with
        different magnetite nano-particle diameters. (b) The stable equilibria, repulsion-attraction transition (trapping), distances at
        $T$=298.15\,K for fixed volume fractions of magnetite ($\Phi=$ 0.005,
        0.01, and 0.015) as function of the fluid permeability calculated with Eq.\,\ref{mu3} for magnetite nanoparticle diameters in the range 2-20\,nm. }
        \label{Figu_trapping}
    \end{figure*}
   
    %According to these studies, 
Since the modeling does not depend on the type of non-magnetic carrier liquid\,\cite{klimchitskaya2019impact,VelichkoKlimchitskaya2020}, it is observed that in general the permeability increases for the fixed volume fraction  of magnetite with increasing the diameter of the magnetite nanoparticles.  
    Using Eq.\,(\ref{mu3}), together with data in Ref.\,\cite{klimchitskaya2019impact,VelichkoKlimchitskaya2020}, we obtain estimates for the static permeability for a range of volume fractions. %from 0 to 0.015 and magnetite diameters from 2 to 20\,nm. 
    Table.\,\ref{MagneticPermability} shows, for varying volume fractions and magnetite particle diameters, a selection of mixed fluid static magnetic permeabilities used in this work.  %\mb{Using mixing models in Lifshitz formula becomes less reliable when separations are similar in size to the nanoparticle size range - I think this can be removed. We can add it if a reviewer brings up the validity of the model}
    
    \begin{table}[h]
        \centering
        \begin{tabular}{|c||c|c||c|c||c|c|}
            \hline
            % after \\: \hline or \cline{col1-col2} \cline{col3-col4} ..
            $\Phi$ & $D$\,(nm) & $\mu_{3}$(0) & $D$\,(nm) & $\mu_{3}$(0) & $D$\,(nm) & $\mu_{3}$(0) \\
            \hline
            \hline
            0.005  & 5         & 1.003        & 10        & 1.024        & 15        & 1.081         \\
            \hline
            0.01   & 5         & 1.006         & 10        & 1.048         & 15        & 1.162        \\
            \hline
            0.015  & 5         & 1.009        & 10        & 1.072         & 15        & 1.243         \\
             \hline
        \end{tabular}
        \caption{\label{MagneticPermability} Static magnetic permeability for
        mixed fluids with volume fraction of magnetite in toluene with $\Phi=$ 0.5\%,\,1\%,\, and 1.5\% %, and 5\% 
        using different diameters ($D$) for the magnetite nano-particles.  
        }
    \end{table}
    \vspace{-0.13in}
Our goal here is to explore whether the usually dominating TM contributions can be overtaken by the TE modes. This can cause sign changes in the product of the reflection coefficients  in Eq.\,\ref{forcedivR} ultimately resulting in a repulsive interaction.
      
The TM-TE dominance balance can lead to surprising results where zero-frequency TE modes can be exploited to tune trapping distances. To achieve this we need systems where the optical properties of the magnetic fluid and the interacting materials are similar. They do not necessarily follow the $\varepsilon_1<\varepsilon_2<\varepsilon_3$ relation required for changing the sign of the force when only dielectric materials are involved \,\cite{Dzya,NinhamParsegianWeiss1970,Richmond_1971}. To understand the magnetic Casimir-Lifshitz force between a polystyrene micron-sized sphere and a Teflon surface in a magnetic fluid in Fig.\,\ref{fig:forcedivRadie_different_different_contributions} we consider the retarded and nonretarded Casimir-Lifshitz force as well as the zero-Matsubara frequency contributions.
 In the non-retarded limit, the interaction is repulsive for all separations. The different retarded models, i.e. including magnetic effects or not, lead to varying crossing over from repulsion to attraction,  trapping, in the submicron range, which is ultimately achieved by the attractive contributions from the zero-frequency TE and TM modes. {A key point is that both the retardation and the full frequency responses are vital to the finding that the zero frequency TE mode has the potential to shift the trapping distance several hundred nanometers. }

In Fig.\,\ref{Figu_trapping}.a we consider materials with fixed optical properties but varying the static permeability for the magnetic fluid. We aim to achieve this by keeping the volume fraction of magnetite particles fixed in the fluid while varying the magnetite particle diameter in the range 5-20\,nm. Here, the magnetic permeability controls the trapping distances, in a very broad range, from submicron distances all the way down to close contact. This is more clearly seen in Fig.\,\ref{Figu_trapping}.b. The trapping occurs when the force turns from repulsion to attraction (shown as the point with zero force in Fig.\,\ref{Figu_trapping}.a) and it depends on the properties of the magnetic fluid. Eq. \ref{mu3} shows that increasing the size of the nanoparticle or equivalently increasing the fraction of the mixture leads to larger $\mu_3$. The greater values for the permeability of the fluid we achieve, the larger the attraction and the smaller the trapping distance. It is interesting to note that nanoparticle trapping in ferrofluids has been realized experimentally by modulating the magnetic susceptibility by applied magnetic fields. Zhang et al.\,\cite{ZhangNature2024} realized from the well-known relation $\mu=1+ \partial M/\partial H$, that changing the external magnetic field leads to tuning of the magnetic susceptibility. By varying $\mu$ with the applied field, one can achieve nanoparticle trapping based on the Casimir effect.  Our results show that instead of an applied magnetic field, Casimir trapping can be achieved using magnetite nanoparticles with appropriate size. 
 For considerations of the magnetic field, as was done in the past, a refined theory is needed that accounts for the influence of the magnetic field on the optical and magnetic properties. 

In contrast to previous research\,\cite{ZhangNature2024}, the proposed setup is unique by the ability to make corrections to the trapping distance that are many orders of magnitude larger by the impact of the nanoparticle size on the transverse electric part of the Casimir force while maintaining the optical properties (i.e., a fixed volume fraction).  We have selected the size range that is most appropriate for measuring Casimir forces in fluids: distances below 100 nm, where the forces are sufficiently strong to produce reliable data\,\cite{ZhangNature2024}.  It also includes the intermediate and longer range that has been in focus for traditional Casimir force measurements\,\cite{Lamo1997,Lamo1998}. Our contribution thus proposes a new and complementary way to measure the effects of the {\it zero-frequency transverse electric mode} and to control particle trapping at both small and large distances.

% \begin{figure}[!h]
%         \centering
%         \includegraphics[width=1\columnwidth]{free_energy.png}
                                                  %         \caption{(Colors online) Force normalized with radius of the PS sphere
%         for a magnetite volume fraction $\Phi=0.01$ and and $T$=298.15\,K with
%         different magnetite nano-particle diameters.}
%         \label{fig:forcedivRadie_different_mu}
%     \end{figure}

%     \begin{figure}[!h]
%         \centering
%         \includegraphics[width=1\columnwidth]{Trapping_distance.png}
%         \caption{(Colors online) The stable equilibria, trapping, distances at
%         $T$=298.15\,K for fixed volume fractions of magnetite ($\Phi=$ 0.005,
%         0.01, and 0.015) as function of the fluid permeability calculated with Eq.\,\ref{mu3} for magnetite nanoparticle diameters in the range 2-20\,nm. 
%         }
%         \label{fig:trapping_distances}
%     \end{figure}

%\textcolor{red}{for the last paragraph - can we say that the magnetic field in Zhang's paper cannot result in great variations in trapping distance and it is only limitted to a small distance range?}

%\textcolor{red}{In the next round, I will make suggested revisions in the Abstarct and last paragraph of the paper; will review the figures, notation and formulas}

%\noindent \textbf{Data availability}:\,The data that support the findings of this study are available in the XXXXXXXXXXXXXXXX

    \section*{Acknowledgements}
    
     LI, and MB's research contributions are part of the project No. 2022/47/P/ST3/01236  co-funded by the National Science Centre and the European Union's Horizon 2020  research and innovation programme under the Marie Sk{\l}odowska-Curie grant agreement No. 945339. SP and UDG acknowledge support from the Marie Sk{\l}odowska-Curie Doctoral Network TIMES, grant No. 101118915 and SPARKLE grant No. 101169225. The research by LI and MB took place at the "ENSEMBLE3-Center of Excellence for nanophononics, advanced materials and novel crystal growth-based    technologies" project (GA No. MAB/2020/14) carried out under the    International Research Agenda programs of the Foundation for Polish Science that are co-financed by the European Union under the European Regional    Development Fund and the European Union Horizon 2020 research and innovation    program Teaming for Excellence (GA. No. 857543) for supporting this work. LI and MB's research contributions to this publication were created as part of the project    of the Minister of Science and Higher Education "Support for the activities of Centers of Excellence established in Poland under the Horizon 2020 program" under contract No. MEiN/2023/DIR/3797. 
    LMW acknowledges support from the US Department of Energy under Award No. DE-FG02-06ER46297. CP research contributions were supported by the European Union's Horizon 2020 research and innovation programme in the project HIYIELD with grant agreement No.~101058694. He acknowledges access to high-performance computing resources via the National Academic Infrastructure for Supercomputing in Sweden
    (NAISS), provided by the National Supercomputer Centre (NSC) at Link\"{o}ping University and by the PDC Center for High Performance Computing at KTH Royal Institute of Technology.

    % \begin{table}[h]
    % \centering
    % \begin{tabular}{c c c}
    %   \hline
    %   \hline
    %    System    &    Approximations \,   &   Power-laws \\
    %     \hline
    %      \makecell{cylinder$||$cylinder \\ (non-conducting)\,\cite{mitchell1973van,langbein1972van,ambrosetti2016wavelike}} & \makecell{Non-retarded \\limit} & $R^{-5}$ \\
    %     \hline
    %        \makecell{cylinder$||$cylinder \\ (conducting)\,\cite{Richmond1972,Davies1973}} & \makecell{Non-retarded \\limit,  $a \ll \lambda_{D}$}  & $R^{-2}$\\
    %     \hline
    %       & \makecell{Non-retarded \\limit,  $a \gg \lambda_{D}$} &  $R^{-2}[\ln({R/a})]^{-3/2}$\\
    %      \hline
    %      \makecell{concentric identical \\ cylinders\,\cite{paresegian2006}} &  \makecell{Non-retarded \\limit}    &  $R^{-2}$\\
    %      \hline
    %     \hline
    % \end{tabular}
    % \caption{\label{SubhojitPowerLawTable} Asymptotic power-law dependency for van der Waals interaction for different cylindrical systems. $a$ is radius of the cylinders and $\lambda_{D}$ is the characteristic Debye length. `$||$' sign denotes the cylinders are parallel to each other.}
    % \end{table}
    % \begin{suppinfo}
    % \label{SI}

    % \end{suppinfo}

    % \onecolumngrid
    \bibliographystyle{apsrev4-2}
   % \bibliography{Subhojitbib}

   %apsrev4-2.bst 2019-01-14 (MD) hand-edited version of apsrev4-1.bst
%Control: key (0)
%Control: author (72) initials jnrlst
%Control: editor formatted (1) identically to author
%Control: production of article title (-1) disabled
%Control: page (0) single
%Control: year (1) truncated
%Control: production of eprint (0) enabled
%

    % \bibliography{GM}

    % \begin{acknowledgments}
    %  J.J.M.,  S.P.,  O.I.M., and M.B.'s research contributions are part of the project No. 2022/47/P/ST3/01236 co-funded by the National Science Centre and the European Union's Horizon 2020 research and innovation programme under the Marie Sk{\l}odowska-Curie grant agreement No. 945339. The research by J.J.M.,  S.P.,  O.I.M., and M.B was carried out at "ENSEMBLE3 - Centre of Excellence for nanophotonics, advanced materials and novel crystal growth-based technologies" project (grant agreement No. MAB/2020/14) carried out within the International Research Agendas programme of the Foundation for Polish Science co-financed by the European Union under the European Regional Development Fund, the European Union's Horizon 2020 research and innovation programme Teaming for Excellence (grant agreement. No. 857543) for support of this work.
    % \end{acknowledgments}

    % \bibliography{Cellulosebib}
\end{document}